\documentclass[epj]{svjour}
\usepackage[utf8]{inputenc}
\usepackage{graphicx,xcolor}
\usepackage{amsmath,amssymb}
\usepackage{multirow}
\usepackage{soul}

\title{A study on the possible merits of using  symptomatic cases to trace the development of the COVID-19 pandemic}
\author{
Gianluca Bonifazi\inst{1,2} \and 
Luca Lista\inst{3, 4} \and
Dario Menasce\inst{5} \and
Mauro Mezzetto\thanks{Corresponding author, e-mail: {\tt mauro.mezzetto@pd.infn.it}.}\inst{,6} \and
Daniele Pedrini\inst{5} \and
Roberto Spighi\inst{2} \and
Antonio Zoccoli\inst{7, 2}
}
\institute{
Universit\`a Politecnica delle Marche \and 
INFN Sezione di Bologna \and
Universit\`a degli Studi di Napoli Federico II \and
INFN Sezione di Napoli \and
INFN Sezione di Milano Bicocca \and
INFN Sezione di Padova \and
Alma Mater Studiorum Universit\`a di Bologna
}

\date{Received: date / Revised version: date}

\abstract{
In a recent work we introduced a novel method to compute the  effective reproduction number $R_t$ and we applied it to describe the development of the COVID-19 outbreak in Italy. The study  is based on the number of daily positive swabs as reported by the Italian Dipartimento di Protezione Civile. Recently, the Italian Istituto Superiore di Sanit\`a made available the data relative of the symptomatic cases, where the reporting date is the date of beginning of symptoms instead of the date of the reporting of the positive swab. In this paper we will discuss merits and drawbacks of this data, quantitatively comparing the quality of the pandemic indicators computed with the two samples.
}

\begin{document}
\maketitle
\section{Introduction}
%
%
The worldwide data about the development of the COVID-19 outbreak is always reported as daily  number of positive swabs, see for instance \cite{JHU}. In a recent paper we introduced a novel method to compute the  effective reproduction number $R_t$ based upon the counting of daily positive swabs \cite{ourpaper}. Positive swabs data suffers from several problems, since  they can be biased by different strategies and response time   for swab data  taken in different regions and  different periods of time.  Data collection is affected by strong weekend effects in  recording the  values, due to reduced capacity of processing swabs on Saturdays and Sundays. Furthermore the reporting of a positive swab introduces a  variable delay between  the date of contagion and   the date of reporting.

Potentially, the reporting of  symptomatic cases, together with the date of  symptom onset, could attenuate most of these issues. In principle, symptomatic cases should suffer less from different strategies of swab data taking, being the most urgent cases to be treated,  and the date of  symptom  onset should be less influenced by weekend effects and should not be affected by additional delays introduced by the processing and reporting of a molecular swab.

On the other hand, the sample of  symptomatic cases is a subset of the total  number of cases,  consequently  the size of the sample is an issue for relatively small populations, like Italian regions or provinces. Furthermore, a bias could be introduced if the true fraction of symptomatic  cases changes during the pandemic because of a modification of the age distribution of infected people. 

From December 6$^{\mathrm{th}}$ 2020, the numbers of symptomatic cases, associated to the date of  symptom onset, are made available in Italy by the daily bulletin of the Istituto Superiore di Sanit\`a (ISS) \cite{ISS}\footnote{It should be noted that  the collection of molecular swabs  began on February 24, 2020 and the reported symptomatic cases reported before this date refer to  positive swabs  only. For this reason the symptomatic cases reported from January 28 to February 24  represent an incomplete sample and we don't consider them in the following for the whole pandemic period.}.  The published data contains the history of all the symptomatic cases on a national basis, while for regions and provinces the daily data  are  only reported.

Data about positive swabs  are  published, since the beginning of the outbreak, by the Italian Dipartimento di Protezione Civile (DPC)\cite{dpc}.
It becomes then possible to compare the information that can be extracted from the full sample of positive swabs with the one from the sub-sample of  symptomatic cases.

In the following, we will work out several indicators to compare the merits and the differences of the two samples.

\section{The data}
\label{sec:data}
\begin{figure}[htbp]
    \centering
    \includegraphics[width=0.70\textwidth]{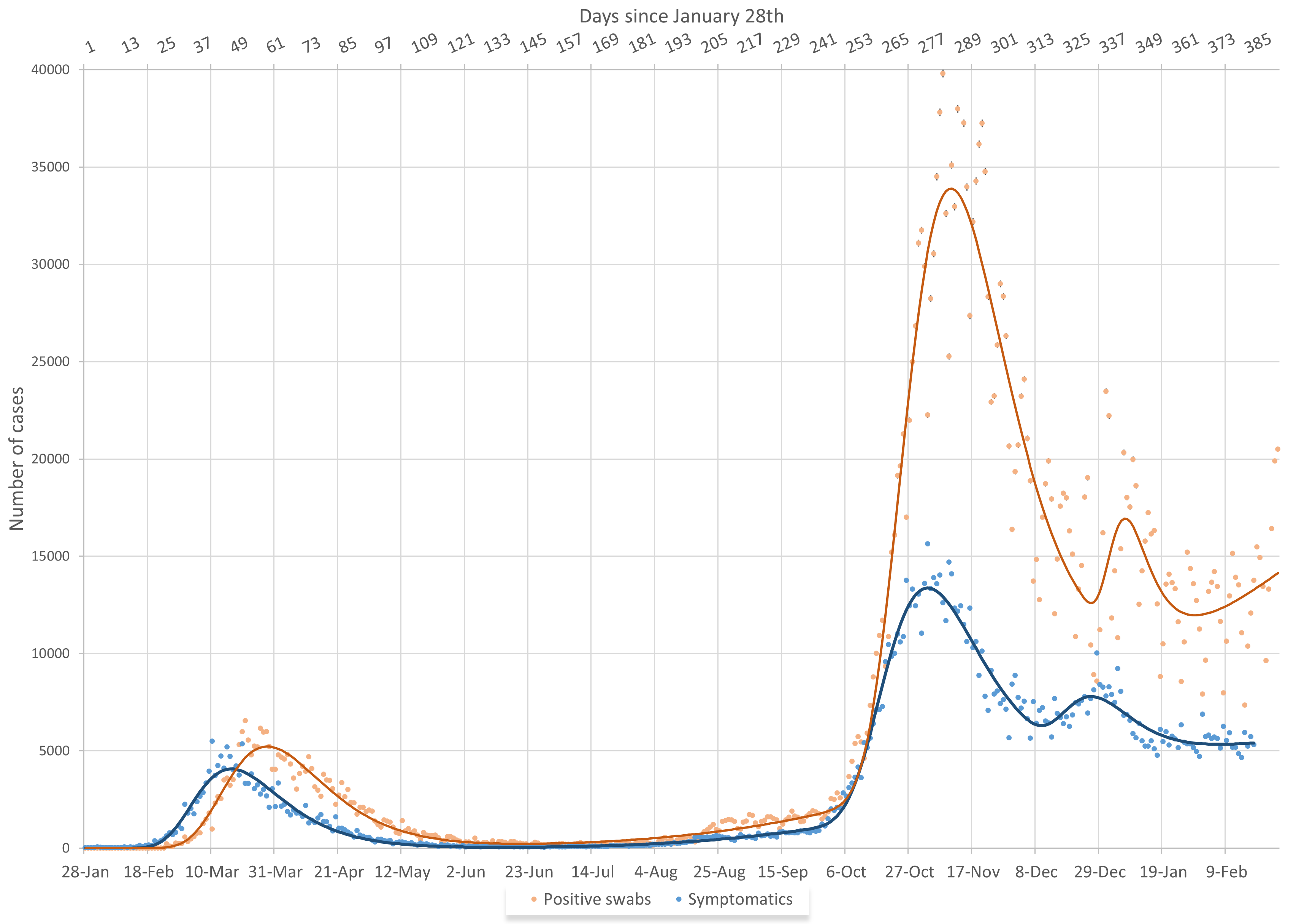}
    \caption{Distribution of the  symptomatic cases (light blue) and of the positive swabs (orange) as reported by ISS and DPC respectively. Poisson errors are drawn but are of the same size of the bullets. The continuous lines are the fits to these data as described in the text.  The upper horizontal axis displays days since January 28$^{\mathrm{th}}$, 2020. }
    \label{fig:data}
\end{figure}

We show in Fig.~\ref{fig:data} the daily data of the  symptomatic cases\footnote{ISS reports two set of data for the symptomatic cases, called ``casi\_inizio\_sintomi" and ``casi\_inizio\_sintomi\_sint". This latter set of data is described as ``number of cases of confirmed SARS-CoV-2 virus infection for which a symptom onset date is indicated except for cases declared as asymptomatic". In absence of any better explanation we use this sample in the following.} and positive swab samples.
We perform a fit to the data with the sum of four derivatives of Gompertz functions, $g(t;a,b,c)$:    
\begin{equation}
    g(t;a,b,c) = a \, e^{-b\, t}e^{-c \, e^{-b \, t}}
    \label{eq:gomp}
\end{equation}
 in order to take care, respectively, of the first phase of the outbreak in the period March--April, the increase of August, the third phase in October--December with the main peak at November and  another  local maxima at the end of December, as reported in the same Fig.~\ref{fig:data}.

The main parameters of the fits are reported in Tab.~\ref{tab:fits}.
\begin{table}[]
    \renewcommand{\arraystretch}{1.8}
    \centering
    \begin{tabular}{ll|ccc}
         & & \begin{tabular}[c]{@{}c@{}}Position\\[-7pt](days)\end{tabular} & \begin{tabular}[c]{@{}c@{}}FWHM\\[-7pt](days)\end{tabular} & \begin{tabular}[c]{@{}c@{}}Pulls\\[-7pt](st. dev.)\end{tabular} \\ \hline
        \multirow{2}{8em}{\textbf{Symptomatics}} & First Peak & $49.4 \pm 0.1$ & $35.0 \pm 0.1 $ & 7.2 \\
         & Second Peak & $279.2 \pm 0.1$ & $51.0 \pm 0.2 $ & 7.1 \\ \hline
        \multirow{2}{8em}{\textbf{Positive swabs}} & First Peak & $61.2 \pm 0.3$ & $39.4 \pm 0.5$ & 8.4 \\
         & Second Peak & $287.1 \pm 0.2$ & $49.9 \pm 0.3 $ & 16.2 
    \end{tabular}
    \caption{For the symptomatics and positive swabs samples we display the fit values of the peak positions expressed in days since January 28$^{\mathrm{th}}$, the values of the Full Width Half Maximum (FWHM) and the standard deviations of the pull distributions for both the first and second peak. The errors of the peak position and of the FWHM are computed according to the covariance matrix of the global fit. Pulls and FWHM are discussed in Sections \ref{sec:pulls} and \ref{sec:fwhm} respectively.}
    \label{tab:fits}
\end{table}

By comparing the dates of the peaks we conclude that the positive swab sample is delayed by about 8  days 8 with respect to the sympotmatics one, which can be considered the average delay between the appearance of the symptoms and the reporting of a positive swab. This number takes into account  that asymptomatic cases are mostly detected by a tracing of the symptomatic cases (which is the cause of delay) and not by a  generic screening of the population.
The fact that the delay at the first peak was larger by 3.9  days with respect to the delay at the second peak could be understood as  more efficient procedures for swab processing developed along the outbreak.

\subsection{Pulls}
\label{sec:pulls}
We can compare the amount of  dispersion present in  the two samples   by computing the pulls of the curves.
Pulls are defined as the difference of the fitted point fit function with the data point, divided by the Poisson error of the data point. Would the Poisson error  be the only source of errors, we would have to find a distribution of pulls with a standard deviation of 1, if the adopted model was the correct one. 

\begin{figure}[htb]
\centering
    \includegraphics[width=0.70\textwidth]{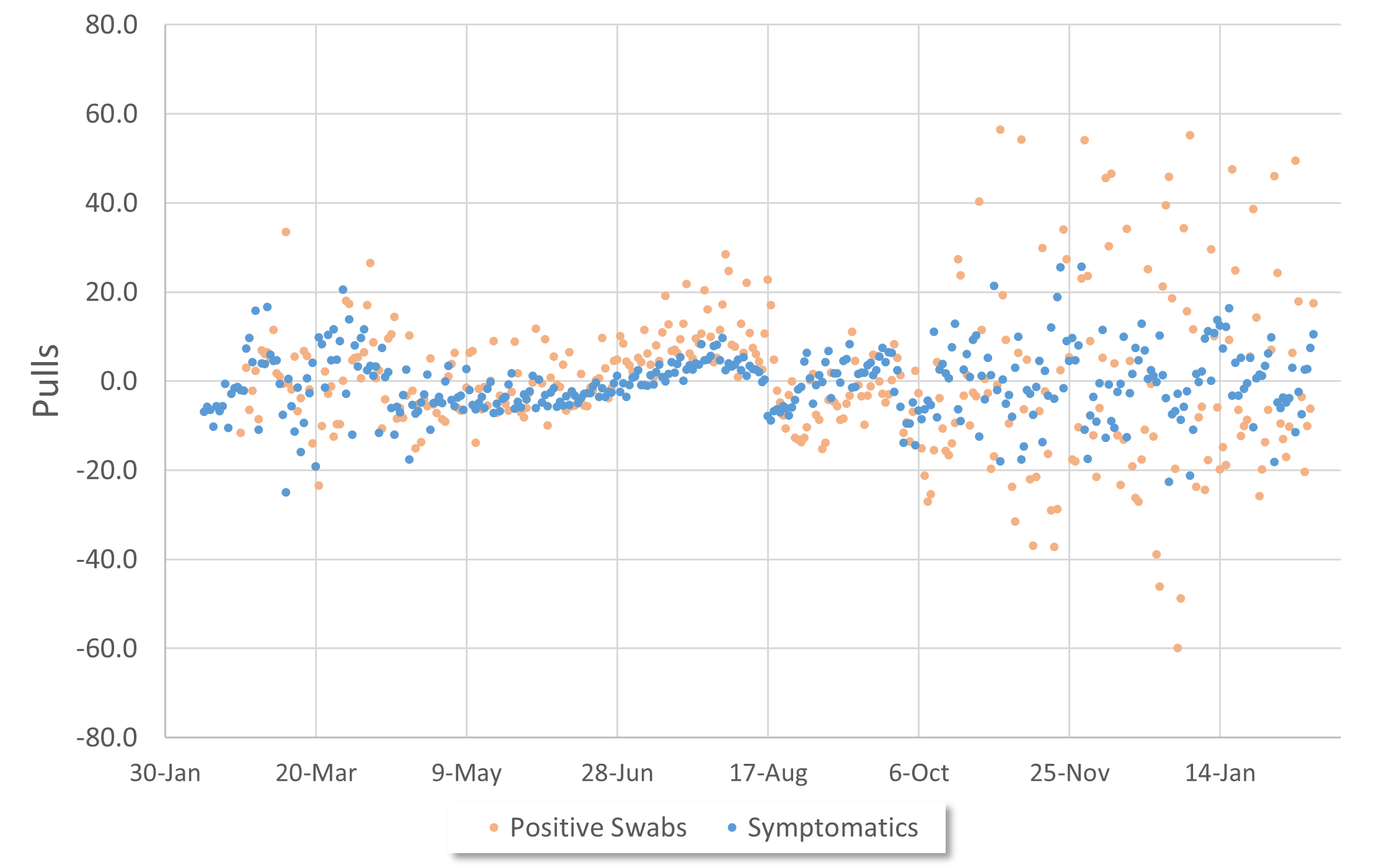}
    \caption{Distribution of the pulls for the four Gompertz fits to the positive swab and symptomatics samples.}
    \label{fig:pulls}
\end{figure}

Pulls distributions are displayed in Fig.~\ref{fig:pulls}.
Limiting the analysis to the first peak, where the  size of the two samples is similar, we obtain a standard deviation of the   residual distributions equal to 7.1 for the  symptomatic cases and of 8.4 for the positive swabs.
A contribution to these high values could be due to a non-perfect parameterization of the data or to an underestimation of the quoted errors that does not take into account additional systematic contributions.

The pulls distribution of the positive swabs  considerably worsen at the second peak.
 We note that in this period antigenic tests initiated to be used by several Italian regions to complement molecular swabs, and from
 January 15$^{\mathrm{th}}$, 2021  antigenic tests are accounted together with the molecular ones. This change of strategy, with the introduction of tests with different efficiency and processing time, could have contributed to the randomization of the data.

We can conclude that the  sample of symptomatic cases does not significantly reduce the dispersion of the data around the central values with respect to the positive swab sample. Changes on the strategy of swab collection can anyway introduce important additional fluctuations in the positive swab sample.

\subsection{Full Width Half Maximum}
\label{sec:fwhm}
Another quantity that could be influenced by additional  fluctuations present in the positive swabs sample is the width of the peaks. If, for instance, the delay between the date of appearance of symptoms and the date of reporting of a positive swab would  follow a broad distribution, this could affect the width of the fitted peaks.

We compute the Full Width Half Maximum (FWHM) of the peaks as the distance between half peak position in the rising and descending part and of the Gompertz curves. According to the values reported in Tab.~\ref{tab:fits}, we found a FWHM of 35.0 and 39.4 days at the first peak and 51.0 and 49.9 days at the second peak for the symptomatic and positive swab samples respectively.  

The symptomatics second wave hardly arrives to half maximum because it restarts for another local maximum; for this reason we don't consider significative the comparison of the widths of the second maximum, where the positive swabs sample nominal value is even smaller.

We consider the differences at the first maximum as an indication of a significant contribution of the dispersion of swab reporting times to the distribution of positive cases. 
We can quantify this contribution,  by considering that in the gaussian approximation the FWHM is equal to 2.35 standard deviations.
If we attribute  the increase of FWHM of the second peak entirely to this effect, the reporting times should have a standard deviation of about 7.7 days.

\subsection{Fraction of  asymptomatic cases}
\begin{figure}[htbp]
    \centering
    \includegraphics[width=0.70\textwidth]{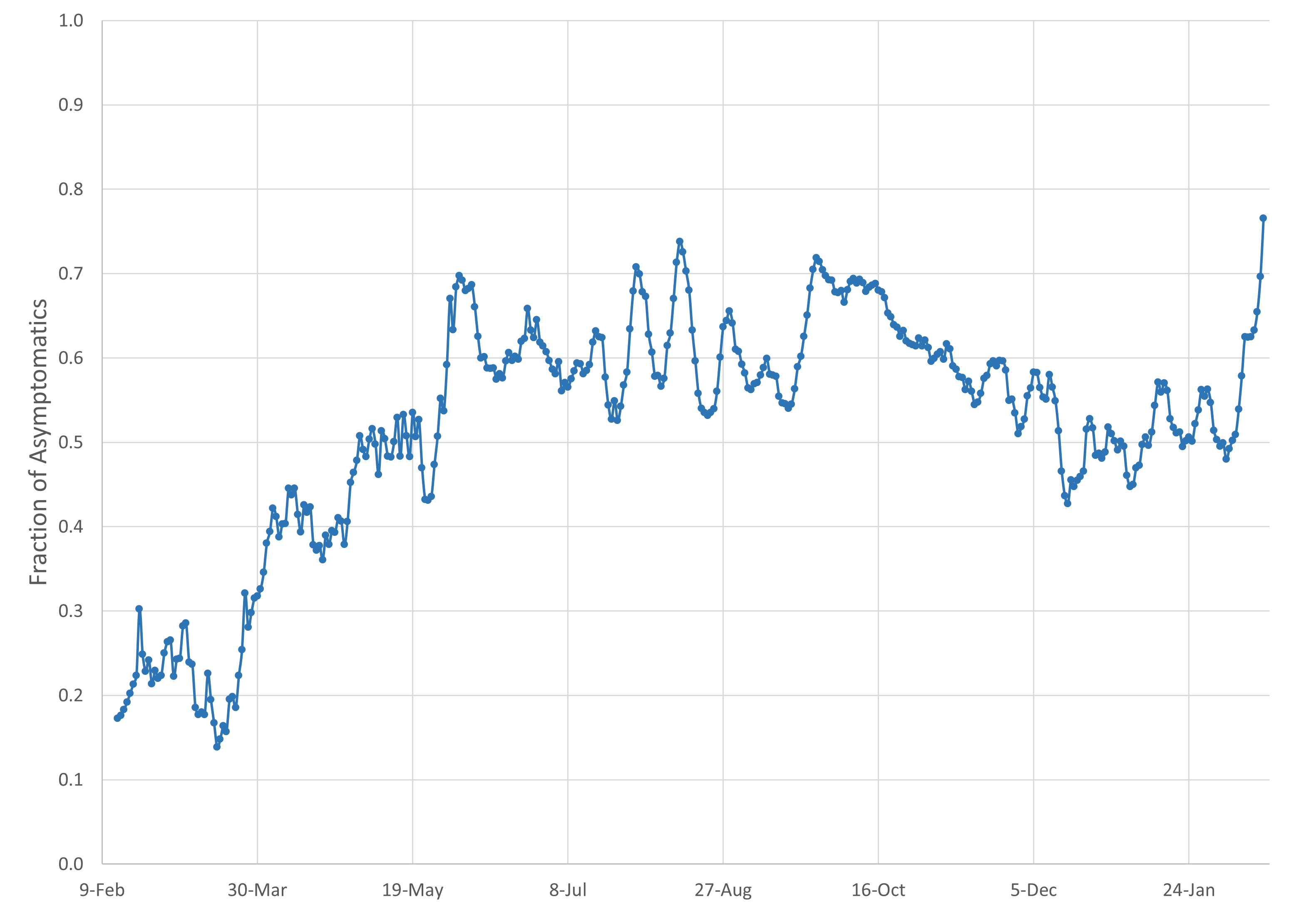}
    \caption{Fraction of  asymptomatic cases in the positive swab sample. Dates are adjusted following those of the symptomatic cases, as discussed in the text.}
    \label{fig:fraction}
\end{figure}

A side result of these comparisons is the distribution of the fraction of asymptomatic cases in the positive swabs sample along the outbreak. If we anticipate by 8 days the positive swabs distribution, according to the above discussion, we can compute day by day the difference of the two data and from this the fraction of asymptomatic cases in the positive swabs sample.
The result is displayed in Fig.~\ref{fig:fraction}. 

%
%
We observe that at the beginning of the pandemic the fraction of  asymptomatic cases  became as small as 0.15 during the first peak of the pandemics. It then grew to about 0.6 at the end of first peak, and remained stable until  the second peak was reached, when  the total number of swabs was probably insufficient to guarantee a proper tracing and the fraction of asymptomatics decreased to 0.5.
\subsection{Stability of the symptomatic sample}
\label{Stability}

\begin{figure}[htbp]
    \centering
    \includegraphics[width=0.70\textwidth]{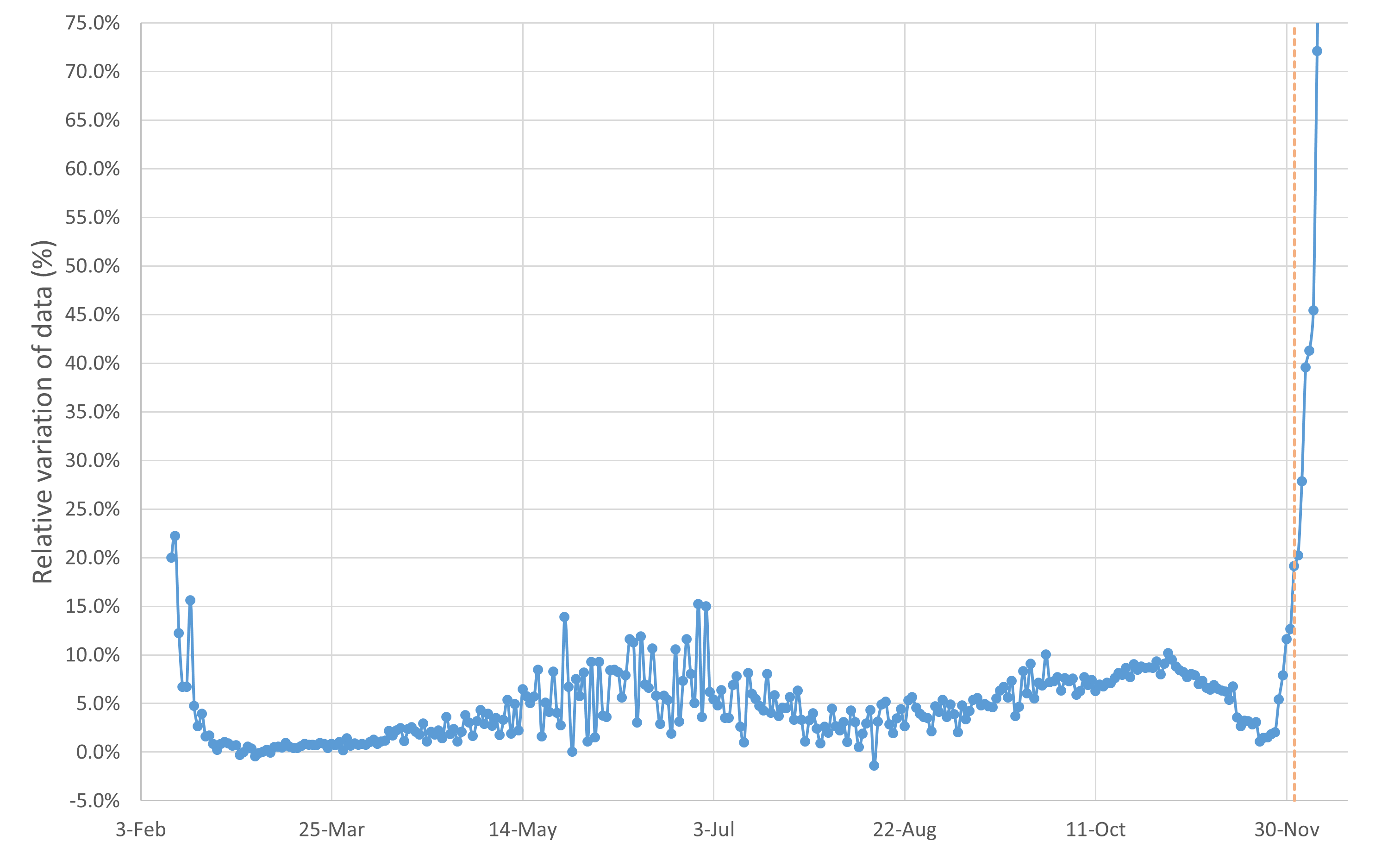}
    \caption{Relative difference, in percentage, of the counts of systematic cases as reported on December 16$^{\mathrm{th}}$, 2020 ($n_i$) and on March 1$^\mathrm{st}$, 2021 ($n'_i$) expressed as $(n'_i-n_i)/n_i$ (\%). The vertical dashed line indicates the 14$^\mathrm{th}$ day to the end of the period. The vertical scale is truncated, so not all the values of the last 14 days are visible.}
    \label{fig:deltat}
\end{figure}

The number of symptomatic cases needs time to stabilize, since they are collected after the reporting of a positive swab and it takes time to execute, process and report a molecular swab after the onset of symptoms.
To check the stability of the sample we considered the data as they were collected on December 16$^{\mathrm{th}}$, 2020 and as they were later updated on March 1$^\mathrm{st}$, 2021;
Fig.~\ref{fig:deltat} shows the variation, day by day, between the original data with respect to its updated values,  up to December 16$^{\mathrm{th}}$, 2020.

ISS does not use the last 14 days of data for the computation of $R_t$, \cite{ISS}, and indeed
from Fig.~\ref{fig:deltat} it's evident that the last 14 days undergo to huge variations,  some additional days are probably needed for a better stabilization of the data. We note however that also all previous days are affected by the recounting of data, with variations as big as 10\%. This means that the symptomatic sample never fully stabilizes, and the derived values of $R_t$ are in this way subject to continuous revisions.

\section{The indicators of the development of the outbreak}
\begin{figure}[htbp]
    \centering
    \includegraphics[width=0.70\textwidth]{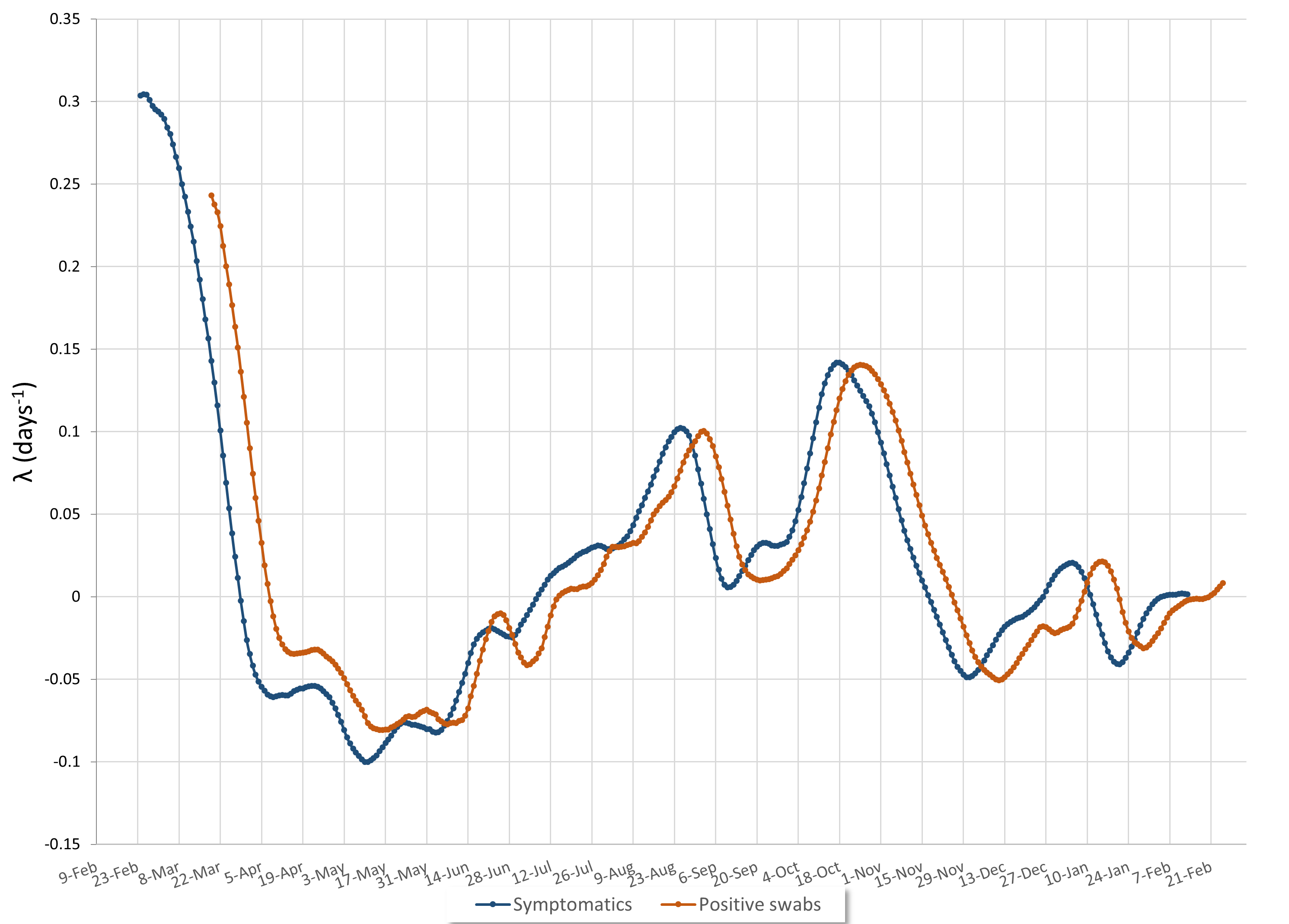}
    \caption{Growth rate $\lambda$  of the exponential fit to reported cases in the last 14 days for the symptomatic sample (blue) and the positive swabs sample (red)}
    \label{fig:t2}
\end{figure}

We discussed in \cite{ourpaper} that the growth rate $\lambda = 1/t_2$, where $t_2$ is the doubling time  of an exponential fit to the data in the last ``$n$'' days, is as good an indicator as $R_t$ for the description of the development of the outbreak. 
$\lambda$ is computed via an exponential fit to the last 14 days  and we display its moving average along 14 days. Results are displayed in Fig.~\ref{fig:t2}. 
The values of $\lambda$ computed with the symptomatics and positive swabs samples have almost the identical behaviour with the characteristic delay of the positive swabs curve.

\begin{figure}[htbp]
    \centering
    \includegraphics[width=0.70\textwidth]{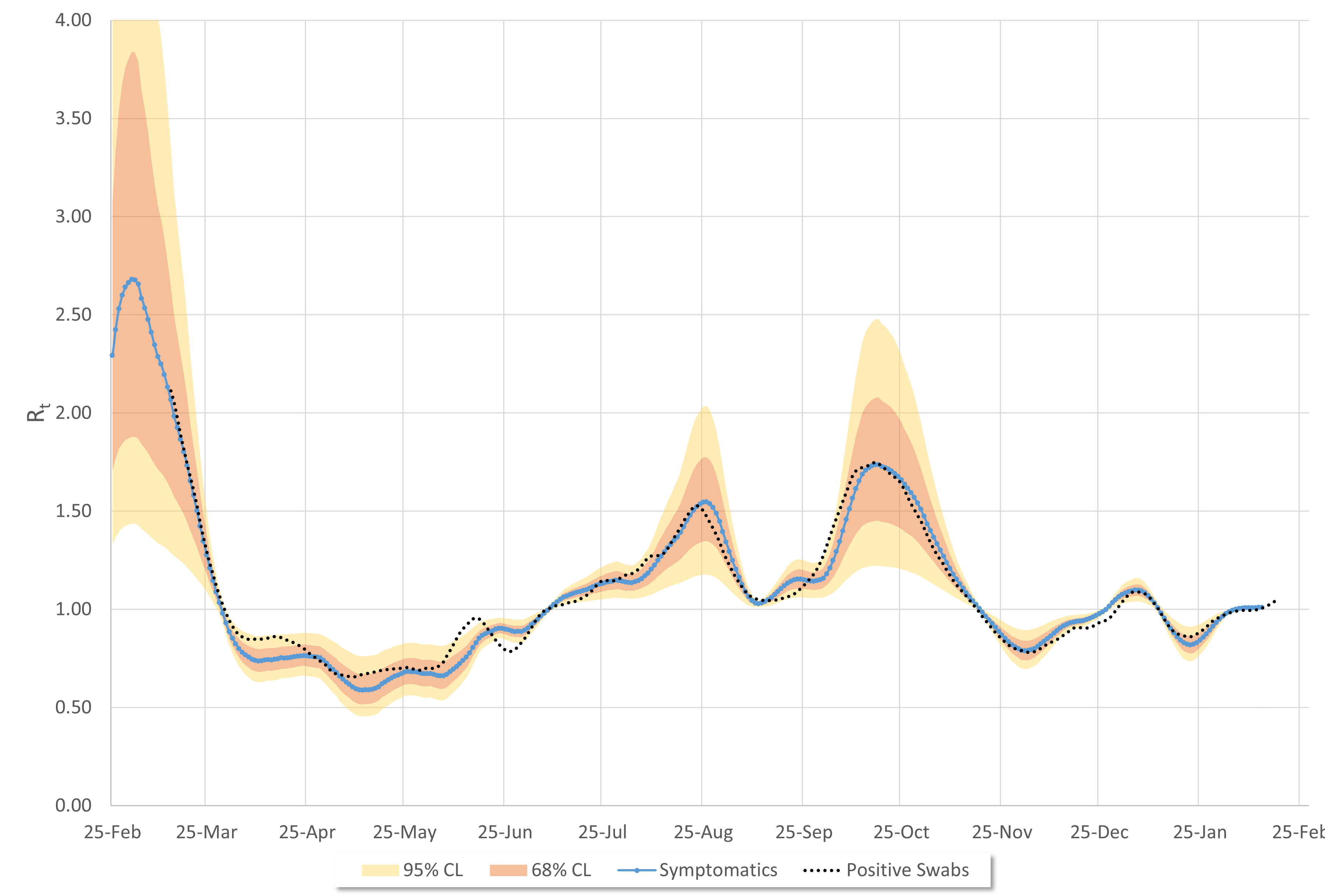}
    \caption{$R_t$ computed for the symptomatic sample (blue) with the uncertainty band up to 68\% confidence level (orange) and up to 95\% confidence level  (yellow). Superimposed is $R_t$ computed with the positive swabs sample (black dotted), moved to the left by 8 days according with the conclusions of the above discussions. $R_t$ is computed with the algorithm published in \cite{ourpaper}. }
    \label{fig:Rt}
\end{figure}

We display the same result in terms of the more familiar $R_t$ in Fig.~\ref{fig:Rt}, computed using the algorithm published in \cite{ourpaper}. In this case, for a better comparison, we anticipate the positive swabs $R_t$  by 8 days, according to the conclusions of Section~\ref{sec:data}.

The two  $R_t$ estimations are  in good agreement within errors, both in shape and absolute values.
The agreement demonstrates that the computation of $R_t$ from the positive swab sample is robust against the strong variations of the fraction of asymptomatics happened during the first peak as well as the significant increase of the dispersion of collected data happened during the second peak.

\begin{figure}[htbp]
    \centering
     \includegraphics[width=0.47\textwidth]{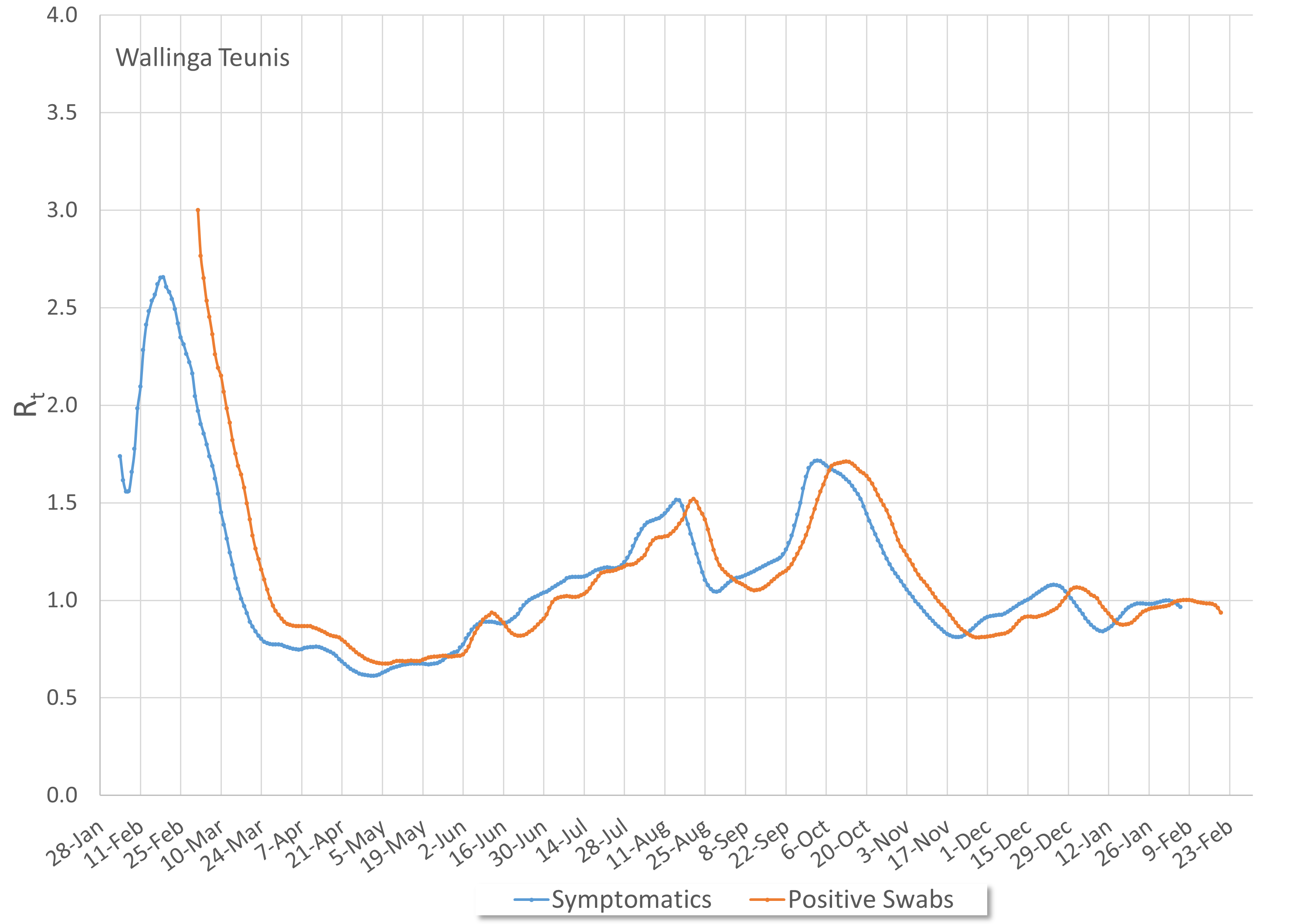}
     \includegraphics[width=0.47\textwidth]{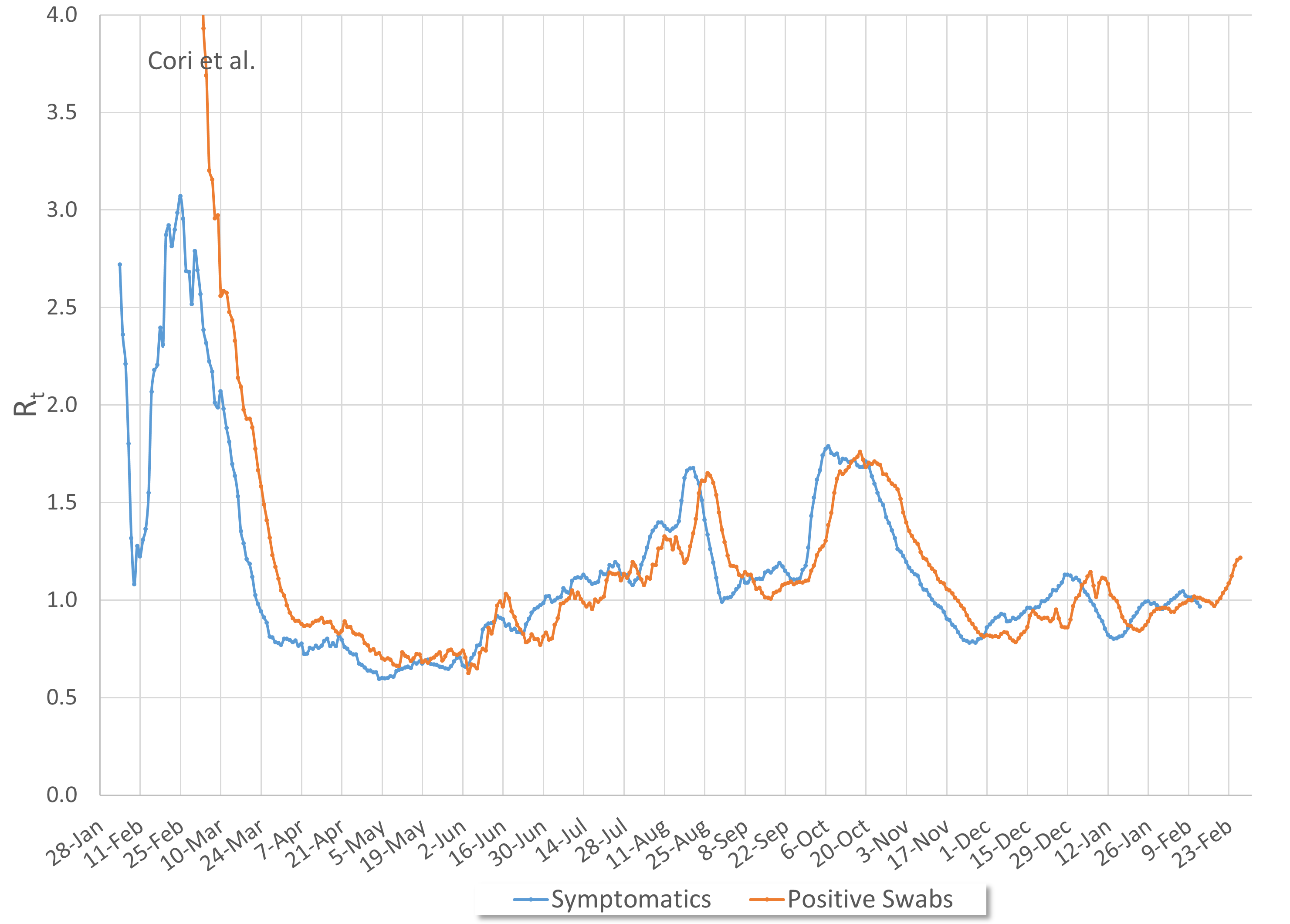}
     \includegraphics[width=0.47\textwidth]{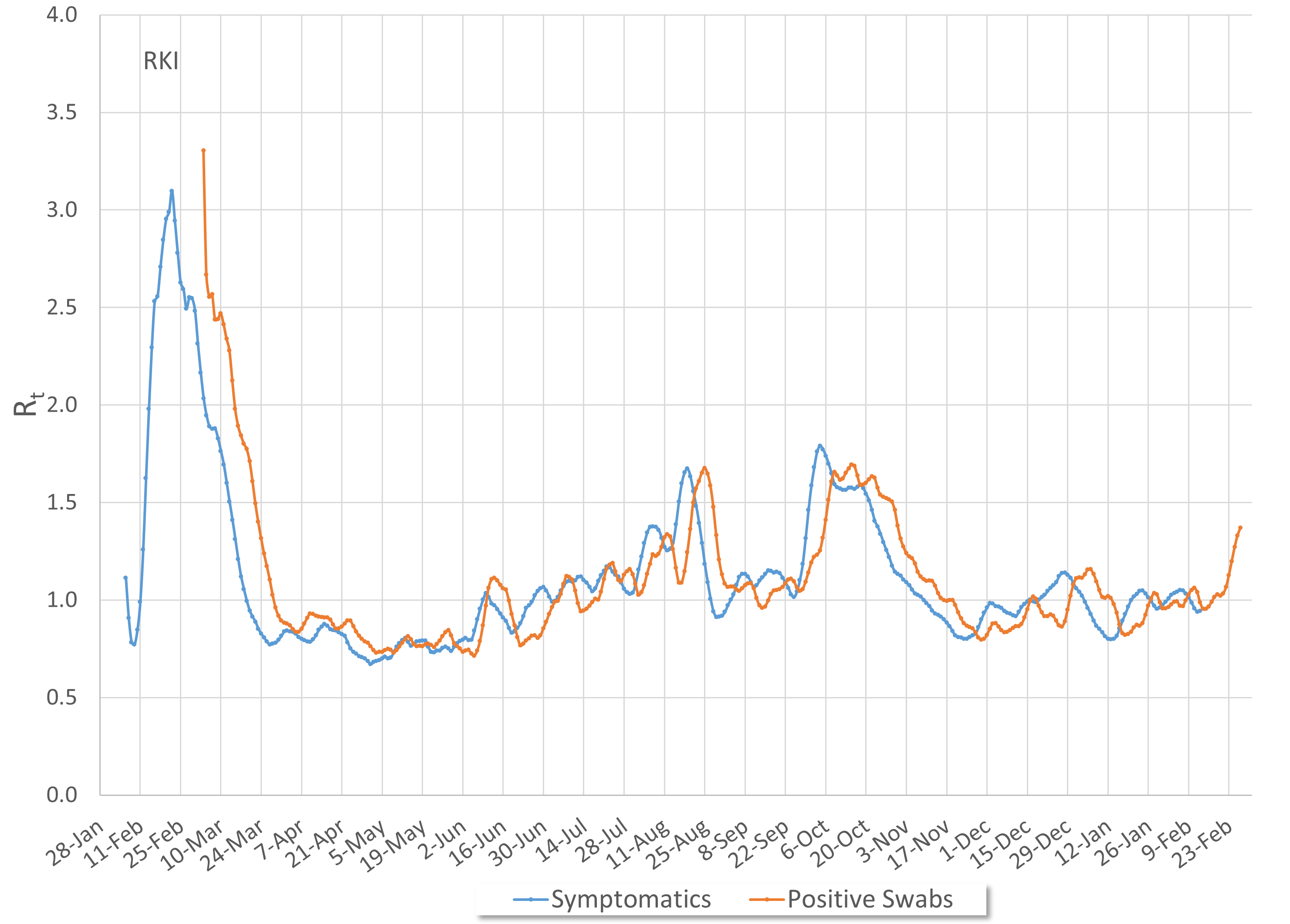}
     \includegraphics[width=0.47\textwidth]{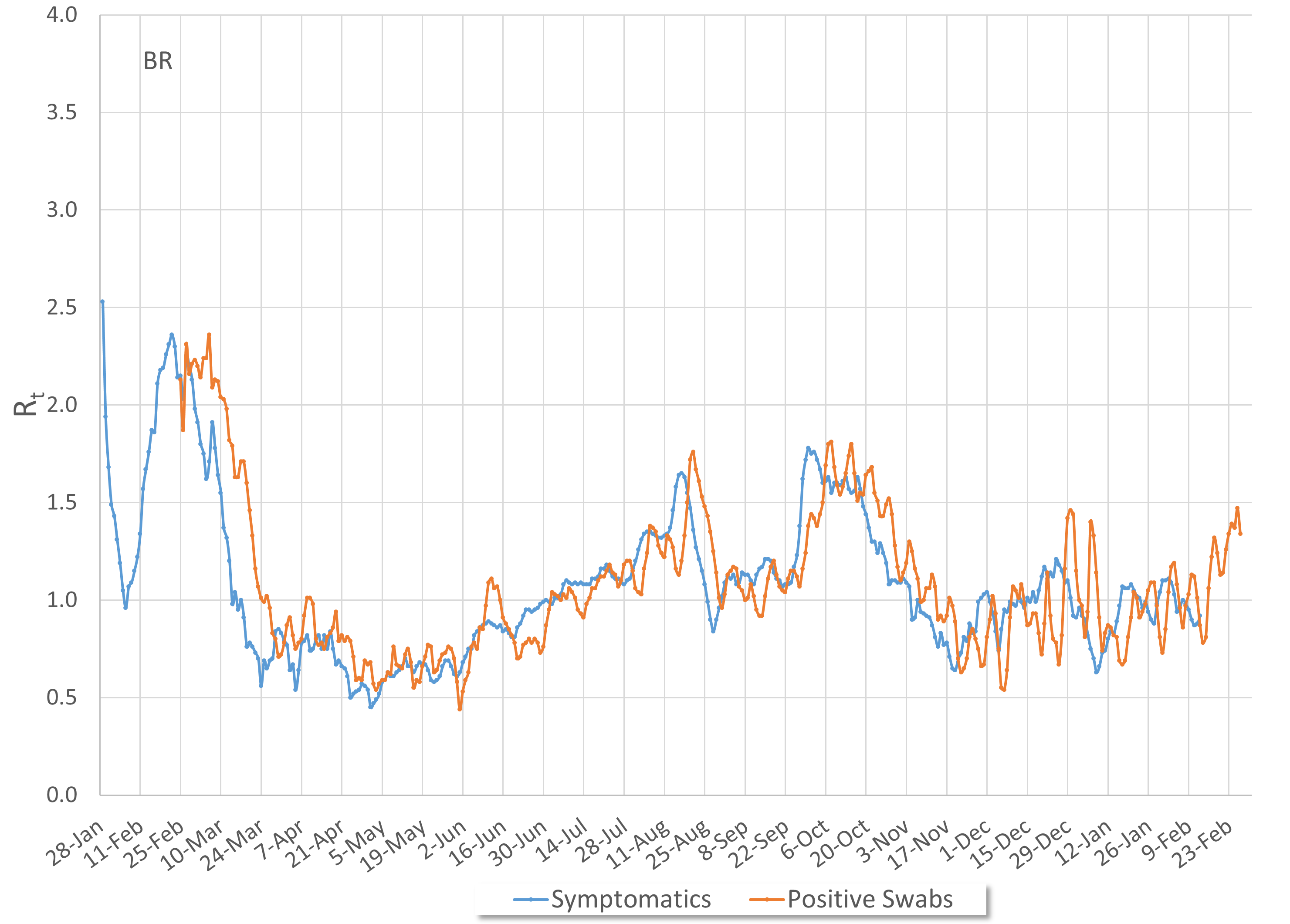}
    \caption{$R_t$ computed for the symptomatic sample (blue) and the positive swabs sample (orange) with four different algorithms (see the text). From upper left,  clockwise:  Wallinga-Teunis, Cori et al., Bettencourt-Ribeiro and RKI.}
    \label{fig:4Rt}
\end{figure}

For the sake of completeness we repeat the same comparison with four of the most common algorithms used in literature to evaluate $R_t$:
Wallinga and Teunis \cite{wallinga}, Cori et al. \cite{cori}, both computed thanks to the public package EpiEstim \cite{epiestim}, Bettencourt-Ribeiro \cite{bettencourt}, computed following the indications of \cite{systrom}, and Robert Koch Institute (RKI) \cite{rki}. The plots are reported in Fig.~\ref{fig:4Rt} and show the identical behaviour of the plot of Fig.~\ref{fig:Rt}. According to ISS \cite{ISS}, their official value of $R_t$ is computed with the Cori et al. algorithm.

\section{Conclusions}
We have compared the information that can be extracted about the development of the COVID-19 outbreak in Italy by using the daily new cases reported for the infected with symptoms  along with the total sample of positive swabs.
The symptomatics  is a  particularly valuable control sample because  it suffers of less systematic effects than the positive swabs sample.

We observe a modest reduction of the dispersion of the data with the  sample of symptomatic cases and a better definition of the peaks in the distribution of the daily positive cases. The differences between the two curves lie mostly in a delay  between the appearance of the symptoms and the date of the reported positive swab,  amounting to about 8 days. By applying this correction, the two samples are  comparable and the corresponding two extracted distributions of $R_t$ turn out to be in good agreement within errors.

One could believe that the  symptomatic sample, being preempted by about 8 days, could identify in advance the trends of the outbreak, however, as discussed in Section~\ref{Stability}, it needs 14 days to be correctly determined.  Considering this effect, the  symptomatic cases sample, as a real-time estimator, is retarded with respect to the positive swabs sample.
The sample of positive swabs provides real-time evaluations of $R_t$ that are faster and more stable.

We conclude that the sample of the positive swabs can be safely used to monitor the development of the COVID-19 outbreak.

We publish  daily estimates of $R_t$ in real time, together with more information about the development of the Italian outbreak in \cite{our web site}. Daily values for the major world countries are also reported.

\section{Acknowledgements}
We acknowledge the effort of ISS of making public the data of the symptomatic cases of COVID-19. The present work has been done in the context of the INFN CovidStat project that produces an analysis of the public Italian COVID-19 data. The results of the analysis are published and updated daily on the website 
{\tt covid19.infn.it/}. The project has been supported in various ways by a number of people from different INFN Units. In particular, we wish to thank, in alphabetic order: Stefano Antonelli (CNAF), Fabio Bredo (Padova Unit), Luca Carbone (Milano-Bicocca Unit), Francesca Cuicchio (Communication Office), Mauro Dinardo (Milano-Bicocca Unit), Paolo Dini (Milano-Bicocca Unit), Rosario Esposito (Naples Unit), Stefano Longo (CNAF), and Stefano Zani (CNAF). We also wish to thank Prof. Domenico Ursino (Universit\`a Politecnica delle Marche) for his supportive contribution.


\begin{thebibliography}{}
\bibitem{JHU}
E. Dong, H. Du, L. Gardner, {\it An interactive web-based dashboard to track COVID-19 in real time.} Lancet Infect. Dis. 20, 533–534 (2020). doi:10.1016/S1473-3099(20)30120-1 Medline

\bibitem{ourpaper}
G. Bonifazi {\it et al.},
{\it A simplified estimate of the Effective Reproduction Number Rt using its relation with the doubling time and application to Italian COVID-19 data},
  arXiv:2012.05194 (2020), submitted to the European Physical Journal Plus.
  
  \bibitem{ISS}
https://www.epicentro.iss.it/coronavirus/sars-cov-2-dashboard

\bibitem{dpc}
Dipartimento della Protezione Civile, {\it Dati COVID-19 Italia},
https://github.com/pcm-dpc/COVID-19


\bibitem{wallinga}
J. Wallinga and P. Teunis, {\it Different Epidemic Curves for Severe Acute Respiratory Syndrome Reveal Similar Impacts of Control Measures}, American Journal of Epidemiology, Volume 160, Issue 6, 15 September 2004, Pages 509–516. https://doi.org/10.1093/aje/kwh255

\bibitem{cori}
A. Cori, N. M. Ferguson, C. Fraser and S. Cauchemez, {\it A New Framework and Software to Estimate Time-Varying Reproduction Numbers During Epidemics}, American Journal of Epidemiology, Volume 178, Issue 9, 1 November 2013, Pages 1505–1512, https://doi.org/10.1093/aje/kwt133

\bibitem{epiestim}
EpiEstim: {\it Estimate Time Varying Reproduction Numbers from Epidemic Curves},
https://cran.r-project.org/web/packages/EpiEstim/index.html


\bibitem{bettencourt}
    L. M. A. Bettencourt and R. M. Ribeiro, 
    {\it Real Time Bayesian Estimation of the Epidemic Potential of Emerging Infectious Diseases},
    PLoS ONE, Volume 3, Issue 5, e2185, 2008,
    https://doi.org/10.1371/journal.pone.0002185

\bibitem{systrom}
K. Systrom, {\it The Metric We Need to Manage COVID-19. 
$R_t$: the effective reproduction number}, 2020,
http://systrom.com/blog/the-metric-we-need-to-manage-covid-19/

    

\bibitem{rki}
Robert Koch Institut, {\it Erläuterung der Schätzung der zeitlich variierenden Reproduktionszahl} (2020),
https://www.rki.de/DE/Content/InfAZ/N/Neuartiges\_Coronavirus/Projekte\_RKI/R-Wert-Erlaeuterung.pdf



\bibitem{our web site}
CovidStat INFN, https://covid19.infn.it/
\end{thebibliography}
\end{document}